\documentclass[12pt]{article}

\newcommand{\be}{\begin{equation}}
\newcommand{\ee}{\end{equation}}
\newcommand{\bi}[1]{\vspace{-3mm} \bibitem{#1}}

\begin{document}
\begin{center}
Journal of Physics A. Vol.39. No.26. (2006) pp.8409-8425.
\end{center}

\begin{center}
{\Large \bf Fractional Variations for Dynamical Systems:}
\vskip 5 mm
{\Large \bf Hamilton and Lagrange Approaches}
\vskip 5 mm

{\large \bf Vasily E. Tarasov }\\

\vskip 3mm

{\it Skobeltsyn Institute of Nuclear Physics, \\
Moscow State University, Moscow 119992, Russia } \\
{E-mail: tarasov@theory.sinp.msu.ru}
\end{center}

\begin{abstract}
Fractional generalization of an
exterior derivative for calculus of variations is defined.
The Hamilton and Lagrange approaches are considered.
Fractional Hamilton and Euler-Lagrange equations are derived.
Fractional equations of motion are obtained by fractional 
variation of Lagrangian and Hamiltonian 
that have only integer derivatives.
\end{abstract}



\section{Introduction}

The theory of derivatives of noninteger order \cite{OS,SKM} goes back 
to Leibniz, Liouville, Riemann, Grunwald and Letnikov \cite{SKM}. 
Derivatives and integrals of fractional order have found many
applications in recent studies in mechanics and physics.
In a fairly short period of time the list of such 
applications becomes long.
For example, it includes chaotic dynamics \cite{Zaslavsky1,Zaslavsky2},
mechanics of fractal media \cite{Mainardi,Hilfer,Nig1,Media},
quantum mechanics \cite{Laskin,Naber}, 
physical kinetics 
\cite{Zaslavsky1,Zaslavsky7,SZ,ZE,Zaslavsky6,Physica2005,TZ2},
plasma physics \cite{CLZ,Plasma2005}, 
astrophysics \cite{CMDA},
long-range dissipation \cite{GM,M}, 
mechanics of non-Hamiltonian systems \cite{nonHam,JPA2005-2},
theory of long-range interaction \cite{Lask,TZ3,KZT}, 
anomalous diffusion and transport theory 
\cite{Zaslavsky1,Montr,Uch} and many others physical topics.

In mathematics and theoretical physics, 
the variational (functional) derivative is a 
generalization of the usual derivative 
that arises in the calculus of variations. 
In a variation instead of 
differentiating a function with respect to a variable, 
one differentiates a functional with respect to a function. 
In this paper, we consider the fractional generalization of 
variational (functional) exterior derivatives. 

The main results are derived in sections 4.2, 5.2, 6.2, and 6.3.
In sections 2, 3, 4.1, 5.1, 6.1, brief reviews of fractional derivatives, 
differential forms, Hamiltonian systems
are considered to fix notation and 
provide convenient references. 
In section 2, a brief review of differential forms is considered.
In section 3, we consider Hamiltonian and fractional Hamiltonian 
systems \cite{JPA2005-2}.
In section 4, we define the fractional variations in Hamilton's approach 
to describe the motion. The fractional generalization of stationary
action principle is suggested. 
In section 5, we discuss the fractional variations in 
Lagrange's approach to describe the motion, 
and the fractional generalization of stationary
action principle is suggested. 
In section 6, we consider the generalization of action principle
to non-Hamiltonian systems. The fractional equations of motion
with friction are discussed.
Finally, a short conclusion is given in section 7.

\section{Fractional Derivatives and Differential Forms}

\subsection{Differential Forms}

In this subsection, a brief review of differential forms \cite{DNF,Grif} 
is considered to fix notation and provide a convenient reference. \\

{\bf Definition 1.}
{\it A differential 1-form 
\be \label{omega} \omega=F^i(x)dx_i \ee
is called an exact 1-form in $R^n$ if the 
vector field $F^i(x)$ can be presented as
\be
F^i(x)=-\frac{\partial V}{\partial x_i},
\ee
where $V=V(x)$ is a continuously differentiable function. } \\

In this case, the differential form (\ref{omega})
is an exact form $\omega=-dV$, where
$V=V(x)$ is a continuously differentiable function (0-form). 
Here $d$ is the exterior derivative \cite{DNF}.
The exterior derivative of the function $V$ is the 1-form 
$dV=dx_i \; {\partial V}/{\partial x_i}$
written in a coordinate chart $(x_1,...,x_n)$.
For the k-form $\omega_k$ and the $l$-form $\omega_l$, 
the exterior derivative obeys the relation 
\be
d (\omega_k \wedge \omega_l)=(d \omega_k) \wedge \omega_l+ (-1)^{k} 
\omega_k \wedge d\omega_l .
\ee
Here $k$ and $l$ are integers.   
Note that $dd \omega= 0$ for any form $\omega$. 
If $d\omega=0$, then  $\omega$ is called a closed form. 

In mathematics \cite{DNF}, 
the concepts of closed and exact forms are defined 
by the equation $d\omega =0$
for a given $\omega$ to be a closed form,
and $\omega=dh$ for an exact form.
It is known that to be exact is a sufficient condition to be closed. 
In abstract terms the question of whether this is also a necessary condition 
is a way of detecting topological information by differential conditions. \\

{\bf Proposition 1.}
{\it If a smooth vector field ${\bf F}={\bf e}_i F^i(x)$ 
satisfies the relations 
\be \label{FxFx} \frac{\partial F^i}{\partial x_j} - 
\frac{\partial F^j}{\partial x_i}=0 \ee
on a contractible open subset X of $R^n$,
then (\ref{omega}) is the exact form such that }
\be \label{ps} \omega=- \frac{\partial V(x)}{\partial x_i} dx_i . \ee

{\bf Proof.} Let us consider the forms (\ref{omega}). 
The formula for the exterior derivative of (\ref{omega}) is
\[ d \omega=\frac{1}{2} \left(\frac{\partial F^i}{\partial x_j} - 
\frac{\partial F^j}{\partial x_i} \right) dx_j \wedge dx_i , \]
where $\wedge$ is the wedge product \cite{DNF}. 
Therefore the condition for $\omega$ to be closed is (\ref{FxFx}).
If $F^i=-{\partial V}/{\partial x_i}$, then
the implication from 'exact' to 'closed' is a consequence 
of the permutability of the second derivatives. 
For the smooth function $V=V(x)$, the second derivative commute, 
and equation (\ref{FxFx}) holds. \\

 
\subsection{Fractional Differential Forms}

If the partial derivatives in the definition of the exterior derivative 
\[ d=dx_i \frac{\partial}{\partial x_i} \]
are allowed to assume fractional order, then 
a fractional exterior derivative is defined \cite{FDF1} by 
\be \label{11a}
d^{\alpha}=(dx_i)^{\alpha} {\bf D}^{\alpha}_{x_i} .\ee 
Here we use 
\be \label{df} {\bf D}^{\alpha}_{x}f(x)=\frac{1}{\Gamma (m-\alpha)}
\int^x_{0} \frac{dy}{(x-y)^{\alpha-m+1}}
\frac{\partial^m f(y)}{\partial y^m}   , \ee
where $\alpha>0$, and $m$ is the first 
whole number greater than  or equal to $\alpha$.
Equation (\ref{df}) defines the Caputo fractional derivatives
\cite{Caputo,Caputo2,Mainardi,Podlubny} of order $\alpha >0$. \\

{\bf Definition 2.} 
{\it A differential 1-form 
\be \label{omega-a} \omega_{\alpha}=F^i(x)(dx_i)^{\alpha} \ee 
is called an exact fractional form
if the the vector field $F^i(x)$ can be represented as
\be \label{FDaV} F^i(x)=-{\bf D}^{\alpha}_{x_i} V ,\ee
where $V=V(x)$ is a continuously differentiable function, and
${\bf D}^{\alpha}_{x_i}$ is a derivative of  order $\alpha$. } \\

Using (\ref{11a}) the exact fractional form can be represented as
\[ \omega_{\alpha}=-d^{\alpha} V=-(dx_i)^{\alpha} {\bf D}^{\alpha}_{x_i}V .\]
Therefore, we have (\ref{FDaV}).

Note that equation (\ref{omega-a}) is a fractional generalization
of the differential form (\ref{omega}). 
Obviously that fractional 1-form $\omega_{\alpha}$
can be closed when the differential 1-form $\omega=\omega_1$ is not closed. 
The fractional analogue of proposition 1 has the form. \\

{\bf Proposition 2.}
{\it If a smooth vector field ${\bf F}={\bf e}_i F^i(x)$ 
on a contractible open subset X of $R^n$ 
satisfies the relations
\be \label{FHC} {\bf D}^{\alpha}_{x_j} F^i- {\bf D}^{\alpha}_{x_i} F^j=0 , \ee
then the form (\ref{omega-a}) is an exact fractional 1-form such that 
\be \label{fps} \omega_{\alpha}=-{\bf D}^{\alpha}_{x_i} V(x) , \ee
where $V(x)$ is a continuous differentiable function and 
${\bf D}^{\alpha}_{x_i}V=-F^i$ }. \\

{\bf Proof}. This proposition is a corollary of the 
fractional generalization of the Poincare lemma \cite{FDF2}.
The Poincare lemma is shown \cite{FDF1,FDF2}
to be true for exterior fractional derivative. \\

Note that we can generalize the definition of 
fractional exterior derivative by the equation
\be
d^{\alpha}=\sum^{n}_{i=1} (dx_i)^{\alpha_i} D^{\alpha_i}_{x_i} ,
\ee
where $\alpha=(\alpha_1,\alpha_2,...,\alpha_n)$,
and consider the fractional differential 1-forms:
\be
\omega_{\alpha}=\sum^{n}_{i=1} \omega_i(x) (dx_i)^{\alpha_i}.
\ee
In this case, we can derive equations with derivatives
of different orders $\alpha_i$.
For simplicity, we suppose that all $\alpha_i=\alpha$.

\section{Hamiltonian Systems}

In this section, a brief review of Hamiltonian systems \cite{DNF}
and fractional Hamiltonian systems \cite{JPA2005-2}
is considered to fix notations and provide a convenient reference.

\subsection{Definition and Properties of Hamiltonian Systems }

Let us use the canonical coordinates 
$(q_1,...,q_n,p_{1},...,p_{n})$ in the phase space $R^{2n}$. 
We consider a dynamical system that is defined by the equations
\be \label{eq1} \frac{dq_i}{dt}=G^i(q,p), 
\quad \frac{dp_i}{dt}=F^i(q,p) . \ee
The definition of Hamiltonian systems can be realized in the 
following form \cite{JPA2005-1,JPA2005-2,Tartmf3}. \\

{\bf Definition 3.} 
{\it The dynamical system (\ref{eq1}) on the phase space $R^{2n}$, 
is called a Hamiltonian system if 
\be \label{beta} \beta=G^idp_i-F^idq_i, \ee
is a closed form, $d \beta=0$.
A dynamical system is called a non-Hamiltonian system 
if (\ref{beta}) is non-closed, $d \beta \not=0$.} \\

The exterior derivative for the phase space is defined as
\be \label{ed}  d= dq_i \frac{\partial}{\partial q_i}+
dp_i \frac{\partial}{\partial p_i} . \ee
Here and later we mean the sum on the repeated index
$i$ from 1 to n. \\

{\bf Proposition 3.} 
{\it If the right-hand sides of equations (\ref{eq1})
satisfy the conditions  
\be \label{HC1} \frac{\partial G^{i}}{\partial p_j}-
\frac{\partial G^{j}}{\partial p_i}= 0, \quad
\frac{\partial G^{j}}{\partial q_i}+
\frac{\partial F^{i}}{\partial p_j}=0, \quad
\frac{\partial F^{i}}{\partial q_j}-
\frac{\partial F^{j}}{\partial q_i}= 0, \ee
then the dynamical system (\ref{eq1}) is a Hamiltonian system.} \\

{\bf Proof}. 
In the canonical coordinates $(q,p)$, the vector fields  
that define the system have the components $(G^i,F^i)$, 
which are used in equation (\ref{eq1}). 
Let us consider the 1-form (\ref{beta})
The exterior derivative of (\ref{beta}) is written by 
\[ d\beta=d(G^idp_i)-d(F^idq_i).  \]
Then
\be \label{6} d\beta= \frac{\partial G^i}{\partial q_j} 
dq_j \wedge dp_i
+\frac{\partial G^i}{\partial p_j} dp_j \wedge dp_i-
\frac{\partial F^i}{\partial q_j} dq_j \wedge dq_i-
\frac{\partial F^i}{\partial p_j} dp_j \wedge dq_i  . \ee
Here $\wedge$ is the wedge product. 
Equation (\ref{6}) can be presented in an equivalent form
\[ d\beta= \left( \frac{\partial G^j}{\partial q_i} 
+\frac{\partial F^i}{\partial p_j}\right) dq_i \wedge dp_j+
\frac{1}{2}\left( \frac{\partial G^j}{\partial p_i}
-\frac{\partial G^i}{\partial p_j} \right)dp_i \wedge dp_j
+\frac{1}{2}\left(\frac{\partial F^i}{\partial q_j}-
\frac{\partial F^j}{\partial q_i}\right) dq_i \wedge dq_j .  \]
Here we use the skew-symmetry of $dq_i \wedge dq_j$ and $dp_i \wedge dp_j$
with respect index $i$ and $j$.
It is obvious that conditions (\ref{HC1}) lead to the equation $d\beta=0$. 
Equations (\ref{HC1}) are called the Helmholtz 
conditions \cite{Helm,Tartmf3,JPA2005-1,JPA2005-2} for the phase space.\\

{\bf Proposition 4}. 
{\it The dynamical system (\ref{eq1}) on the phase space $R^{2n}$, 
is a Hamiltonian system that is defined by the Hamiltonian $H=H(q,p)$
if the form (\ref{beta}) is an exact form  $\beta=dH$, 
where $H=H(q,p)$ is a continuous differentiable unique 
function on the phase space. } \\

{\bf Proof.} 
Suppose that the form (\ref{beta}) is
\[ \beta=dH=\frac{\partial H}{\partial p_i} dp_i+
\frac{\partial H}{\partial q_i} dq_i. \]
Then the vector fields $(G^i,F^i)$ are
\be \label{20} G^i(q,p)=\frac{\partial H}{\partial p_i}, \quad
F^i(q,p)=-\frac{\partial H}{\partial q_i} . \ee
If $H=H(q,p)$ is a continuous differentiable function,
then the conditions (\ref{HC1}) are satisfied, and
(\ref{eq1}) is a Hamiltonian system.
Substitution of (\ref{20}) into (\ref{eq1}) gives
\be \label{eq2} \frac{dq_i}{dt}=\frac{\partial H}{\partial p_i}, 
\quad \frac{dp_i}{dt}=-\frac{\partial H}{\partial q_i} . \ee
As the result, the equations of motion are uniquely 
defined by the Hamiltonian $H$. \\

\subsection{Fractional Hamiltonian Systems}

Fractional generalization of Hamiltonian systems has been suggested
in \cite{JPA2005-2}.
Fractional analogue of the form (\ref{beta}) can be defined by
\be \label{beta-a} \beta_{\alpha}=G^i (dp_i)^{\alpha}-F^i (dq_i)^{\alpha}. \ee
Let us consider the equations of motion 
\be \label{feq1} A_tq_i=G^i(q,p), 
\quad B_t p_i=F^i(q,p) , \ee
where $A_t$ and $B_t$ are the linear (or nonlinear) operators
that have derivatives (integer or fractional order) 
with respect to time.
As the simple examples, we can consider 
the total time derivatives $A_t=B_t=d/dt$ and equation (\ref{eq1}). 
For the fractional derivatives $A_t=B_t=D^{\alpha}_t$, 
\be \label{feq1b} D^{\alpha}_t q_i=G^i(q,p), 
\quad D^{\alpha}_tp_i=F^i(q,p) . \ee

{\bf Definition 4.} 
{\it The dynamical system (\ref{feq1}) on the phase space $R^{2n}$ 
is called a fractional Hamiltonian system if (\ref{beta-a})
is a closed fractional form 
\be \label{ba} d^{\alpha} \beta_{\alpha}=0 , \ee
where $d^{\alpha}$ is the fractional exterior derivative. 
The system is called a fractional non-Hamiltonian system 
if (\ref{beta-a}) is non-closed fractional form, 
i.e.,  $d^{\alpha} \beta_{\alpha} \not=0$.} \\

The fractional exterior derivative for the phase space $R^{2n}$
is defined as
\be \label{fed2}  d^{\alpha}=(dq_i)^{\alpha} 
{\bf D}^{\alpha}_{q_i}+(dp_i)^{\alpha} {\bf D}^{\alpha}_{p_i},
 \quad \alpha >0 . \ee
Let us consider a fractional generalization of the Helmholtz conditions. \\

{\bf Proposition 5.} 
{\it If the right-hand sides of equations (\ref{feq1})
satisfy the conditions 
\be {\bf D}^{\alpha}_{p_j} G^i-
{\bf D}^{\alpha}_{p_i} G^j= 0, \quad
{\bf D}^{\alpha}_{q_i} G^j+
{\bf D}^{\alpha}_{p_j} F^i=0, \quad
{\bf D}^{\alpha}_{q_j} F^i-
{\bf D}^{\alpha}_{q_i} F^j= 0, \ee
then dynamical system (\ref{feq1}) is a 
fractional Hamiltonian system.} \\


{\bf Proof.}  This proposition has been proved in Ref. \cite{JPA2005-2}.\\

{\bf Proposition 6.} 
{\it The dynamical system (\ref{feq1}) on the phase space  $R^{2n}$ 
is a fractional Hamiltonian system with the Hamiltonian $H=H(q,p)$ 
if (\ref{beta-a}) is an exact fractional form
\be \label{b2a}  \beta_{\alpha}=d^{\alpha} H , \ee
where $H=H(q,p)$ is a continuous 
differentiable function on the phase space. }\\

{\bf Proof}. 
Suppose that the fractional form (\ref{beta-a}) is
\[ \beta_{\alpha}=d^{\alpha}H=(dp_i)^{\alpha}
{\bf D}^{\alpha}_{p_i} H +(dq_i)^{\alpha} {\bf D}^{\alpha}_{q_i} H . \]
Then
\[ G^i(q,p)={\bf D}^{\alpha}_{p_i} H, \quad
F^i(q,p)=-{\bf D}^{\alpha}_{q_i} H , \]
and equation (\ref{feq1}) gives
\be \label{feq2} A_t q_i={\bf D}^{\alpha}_{p_i} H, 
\quad B_t p_i=-{\bf D}^{\alpha}_{q_i} H . \ee
These equations describe the motion of fractional Hamiltonian systems.


\section{Hamilton's Approach}

\subsection{Hamilton's Equations of Integer Order}

Let us consider Hamiltonian systems in the extended phase 
space $M^{2n+1}=R^1 \times R^n\times R^n$ of coordinates $(t,q,p)$.
The motion of systems is defined by 
stationary states of the action functional
\be \label{Sqp}
S[q,p]=\int [p \dot{q}-H(t,q,p)]  dt,
\ee
where $H$ is a Hamiltonian of the system, $\dot{q}=dq/dt$, and
both $q$ and $p$ are assumed to be independent functions of time.
In classical mechanics, the trajectory of an object is derived 
by finding the path for which the action integral (\ref{Sqp}) 
is stationary (a minimum or a saddle point). 

In Hamilton's approach the action functional  (\ref{Sqp})
can be written as 
\be
S[q,p]=\int \omega_h ,
\ee
where
\be \label{CP}
\omega_h=pdq-Hdt .
\ee
The form (\ref{CP}) is called the {\it Poincare-Cartan 1-form} or 
the {\it action 1-form}.

The Poincare-Cartan form looks like the integrand of 
the action or the  Lagrangian.  However, it is a differential  form 
on the extended phase space $M^{2n+1}$ of $(t,q,p)$, not  a function.  
Once we integrate it over a curve $C$ in $M^{2n+1}$, 
we get the action:
\be
S[q,p]=\int_C \omega_h =\int^B_A [p dq -H(t,q,p)dt] .
\ee
The integration is taken from $A$ to $B$ in the extended phase 
space $M^{2n+1}$. 

Now suppose we integrate from $A$ to $B$ along two slightly different 
paths and take the difference to get a close loop integral.  
To evaluate this integral we can use Stokes' theorem \cite{DNF}.  
In  the  language  of  differential  forms,  Stokes' 
theorem  is  written as
\be \label{Stokes}
\int_{\partial M}\omega=\int_M d \omega .
\ee 
Here, $M$ is a n-dimensional compact orientable manifold with 
boundary $\partial M$ and  $\omega$ is a $(n-1)$-form; 
$'d'$ is its exterior derivative.  
Note that $M$ can be a submanifold of a larger space, 
so that Stokes' theorem actually implies a whole set of relations including 
the familiar Gauss and Stokes laws of ordinary vector calculus. 

Applying equation (\ref{Stokes}) to the difference of 
actions computed along two neighboring paths with $(q,t)$ 
fixed at the endpoints, we get 
\be
\delta S[q,p]=\int_{\sigma} d \omega_h =\int_{\sigma} 
\left(dp \wedge dq-dH \wedge dt \right) ,
\ee
where $\sigma$ denotes the surface area in the extended phase 
space bounded by the two paths from $A$ to $B$.  

The principle of stationary action states that $\delta S=0$ 
for small variations about the true path, with  $(q,t)$ 
fixed at the end points.  
This will be true for arbitrary small variations, if and only 
if $d\omega_h= 0$ for the tangent vector along the extremal path.  

We can consider the exterior derivative of Poincare-Cartan 1-form,
and derive the equations of motion from the condition $d \omega_h=0$. 
Using this condition, we get the Hamilton's equations of motion.
This condition is equivalent to the 
stationary action principle $\delta S[q,p]=0$. \\

{\bf Proposition 7.}\\
{\it The exterior derivative of the Poincare-Cartan 1-form (\ref{CP}) 
is defined by the equation  
\be \label{dCP3}
d\omega_h=\left[ D_t p+D_q H \right]  dt \wedge dq-
\left[ dq-D_p H dt \right] \wedge dp ,
\ee
where }
\be 
D_t =\frac{\partial }{\partial t}, \quad
D_q =\frac{\partial }{\partial q}, \quad
D_p =\frac{\partial }{\partial p} . 
\ee

\vskip 3mm

{\bf Proof}. 
The exterior derivative of the form (\ref{CP})
can be calculated from the equation
\[ d\omega_h=d(pdq)-d(Hdt)=
D_t p dt \wedge dq+D_q p dq \wedge dq+D_p p dp \wedge dq - \]
\be \label{dCP1}
-D_t H dt \wedge dt- D_q H dq \wedge dt-D_p H dp \wedge dt.
\ee
Using $dt \wedge dt=0$, $dp \wedge dt=-dt \wedge dp$, 
and $D_q p=0$, we get
\be \label{dCP2}
d\omega_h=\left[ D_t p+D_q H \right]  dt \wedge dq-
\left[ D_p p dq-D_p H dt \right] \wedge dp . 
\ee
The relation $D_p p=1$ gives
\be \label{dCP3e}
d\omega_h=\left[ D_t p+D_q H \right]  dt \wedge dq-
\left[ dq-D_p H dt \right] \wedge dp .
\ee

\vskip 3mm

{\bf Stationary Action Principle in Hamilton's Approach}  \\
{\it The trajectory of a Hamiltonian system can be derived 
by finding the path for which the Poincare-Cartan 1-form $\omega_{h}$ 
is closed, i.e.,} 
\be \label{SAP}
d \omega_{h}=0 .
\ee

Using the stationary action principle (\ref{SAP}),
we get the equations of motion
\be 
dq-D_p H dt=0, \quad D_t p=-D_q H.
\ee

As the result, we obtain 
\be 
\frac{dq}{dt}=D_p H , \quad \frac{dp}{dt}=-D_q H, 
\ee
which are the well-known Hamilton's equations.

\subsection{Fractional Hamilton's Equations}

The fractional generalization of the form (\ref{CP}) can be defined by
\be \label{fCP}
\omega_{h,\alpha}=p(dq)^{\alpha}-H(dt)^{\alpha} .
\ee
Note that $\omega_{h,\alpha}$ is a fractional 1-form  
that can be called a {\it fractional Poincare-Cartan 1-form}
or {\it fractional action 1-form}.

We can consider the fractional exterior derivative of the form (\ref{fCP}),
and use $d^{\alpha} \omega_{h,\alpha}=0$ to obtain 
the fractional equations of motion. \\

{\bf Proposition 8.}
{\it The fractional exterior derivative of 
the fractional form (\ref{fCP}) is} \\
\be \label{dfCP3}
d^{\alpha}\omega_{h, \alpha}=\left[ D^{\alpha}_t p+D^{\alpha}_q H \right]  
(dt)^{\alpha} \wedge (dq)^{\alpha}-
\left[ \frac{p^{1-\alpha}}{\Gamma(2-\alpha)} (dq)^{\alpha}-
D^{\alpha}_p H (dt)^{\alpha} \right] \wedge (dp)^{\alpha} .
\ee

\vskip 3mm

{\bf Proof}. 
The fractional exterior derivative of the form (\ref{fCP})
is calculated by using the rule 
\[ {\bf D}^{\alpha}_{x} (fg) =\sum^{\infty}_{k=0} 
\left(^{\alpha}_k \right) ({\bf D}^{\alpha-k}_x f )
\frac{\partial^k g}{\partial x^k} ,\]
and the relation
\[ \frac{\partial^k}{\partial x^k} 
\left[(dx)^{\alpha}\right]=0 \quad (k\ge 1) . \]
For example, we have 
\[ d^{\alpha} \left[ A^i (dx_i)^{\alpha} \right]=\sum^{\infty}_{k=0}
(dx_j)^{\alpha} \wedge 
\left(^{\alpha}_k \right)
({\bf D}^{\alpha-k}_{x_j} A^i) 
\frac{\partial^k}{\partial x^k_j} (dx_i)^{\alpha}=\]
\[ =(dx_j)^{\alpha} \wedge (dx_i)^{\alpha}
\left(^{\alpha}_0 \right)
({\bf D}^{\alpha}_{x_j} A^i)=
\left( {\bf D}^{\alpha}_{x_j} A^i \right)
(dx_j)^{\alpha} \wedge (dx_i)^{\alpha} , \]
where 
\[ \left(^{\alpha}_k \right)=
\frac{(-1)^{k-1} \alpha \Gamma(k-\alpha)}{\Gamma(1-\alpha) \Gamma(k+1)}. \]

As the result, 
\[ d^{\alpha}\omega_{h,\alpha}=
d^{\alpha}(p(dq)^{\alpha})-d^{\alpha}(H(dt)^{\alpha})=\]
\[ =(D^{\alpha}_t p) (dt)^{\alpha} \wedge (dq)^{\alpha}+
(D^{\alpha}_q p) (dq)^{\alpha} \wedge (dq)^{\alpha}+
(D^{\alpha}_p p) (dp)^{\alpha} \wedge (dq)^{\alpha}- \]
\be \label{dfCP1} -
(D^{\alpha}_t H) (dt)^{\alpha} \wedge (dt)^{\alpha}- 
(D^{\alpha}_q H) (dq)^{\alpha} \wedge (dt)^{\alpha}-
(D^{\alpha}_p H) (dp)^{\alpha} \wedge (dt)^{\alpha}.
\ee
Here $D^{\alpha}$ are the Riesz or Caputo fractional derivatives
\cite{Caputo,Caputo2,Mainardi,Podlubny}.
Note the Riemann-Liouville fractional derivative leads us 
to dependence of independent coordinates \cite{nonHam,JPA2005-2}:
\be 
D^{\alpha}_q p=p D^{\alpha}_q1 \not=0 . 
\ee
Therefore the fractional equations are more complicated for
Riemann-Liouville derivatives. 
Using $(dt)^{\alpha} \wedge (dt)^{\alpha}=0$, 
$(dp)^{\alpha} \wedge (dt)^{\alpha}=-(dt)^{\alpha} \wedge (dp)^{\alpha}$
and $D^{\alpha}_q p=0$ for Riesz and Caputo derivatives, 
we can rewrite equation (\ref{dfCP1}) in the form
\be \label{dfCP2}
d^{\alpha}\omega_{h,\alpha}=
\left[ D^{\alpha}_t p+ D^{\alpha}_q H \right] 
(dt)^{\alpha} \wedge (dq)^{\alpha}-
\left[ (D^{\alpha}_p p) (dq)^{\alpha}- (D^{\alpha}_p H) (dt)^{\alpha} 
\right] \wedge (dp)^{\alpha}.
\ee
Substitution of 
\be
D^{\alpha}_p p=\frac{p^{1-\alpha}}{\Gamma(2-\alpha)},
\ee
into equation (\ref{dfCP2}) gives (\ref{dfCP3}). \\

{\bf Fractional Action Principle in Hamilton's Approach}  \\
{\it The trajectory of a dynamical system can be derived 
by finding the path for which the 
fractional Poincare-Cartan 1-form $\omega_{h,\alpha}$ is 
fractional closed form, i.e., }
\be \label{FSAP}
d^{\alpha} \omega_{h, \alpha}=0 .
\ee

Here, we consider only fractional Hamiltonian systems. 
The non-Hamiltonian systems are considered in section 6.

Using (\ref{dfCP3}) and (\ref{FSAP}), we get 
\be 
\frac{p^{1-\alpha}}{\Gamma(2-\alpha)} (dq)^{\alpha}-
D^{\alpha}_p H (dt)^{\alpha}=0, \quad 
D^{\alpha}_t p=-D^{\alpha}_q H .
\ee
As the result, we obtain
\be \label{fHe}
\left(\frac{d q}{dt}\right)^{\alpha}=\Gamma(2-\alpha)
p^{\alpha-1} D^{\alpha}_p H ,\quad
D^{\alpha}_t p=-D^{\alpha}_q H .
\ee
These equations are the fractional generalization
of Hamilton's equations.  

For the fractional Poincare-Cartan 1-form
\be \label{fCPb}
\omega_{h,\alpha}=p^{\beta}(dq)^{\alpha}-H(dt)^{\alpha} ,
\ee
equations (\ref{fHe}) are
\be \label{Hb}
\left(\frac{d q}{dt}\right)^{\alpha}=
\frac{\Gamma(\beta+1-\alpha)}{\Gamma(\beta+1)}
p^{\alpha-\beta} D^{\alpha}_p H ,\quad
D^{\alpha}_t p^{\beta}=-D^{\alpha}_q H .
\ee
Note that we cannot use the rule of differentiating 
a composite functions for fractional derivative 
$D^{\alpha}_t p^{\beta}$. Therefore equations (\ref{Hb}) with
$\beta \not= 1$ are more complicated than equation (\ref{fHe}). \\

\section{Lagrange's Approach}

\subsection{Lagrange's Equations of Integer Order}

Suppose $L(t,q,v)$ is a Lagrangian of dynamical system,
where $q$ is coordinate, and $v$ is the velocity.
Let us consider the variational problem on 
the extremum of the action functional
\be
S_0[q,v]=\int L(t,q,v) dt,
\ee
under the additional condition
\be
\dot{q}=v,
\ee
where both $q$ and $v$ are assumed to be independent functions of time.
In this case, $p$ play the role of independent Lagrange multipliers.
Obviously, the indicated variational problem is equivalent
to the problem on the extremum of the action,
\be
S[q,v,p]=\int [L(t,q,v)+p(\dot{q}-v)] dt,
\ee
where already all the variables $q$, $v$, $p$ have to be varied.
The corresponding Lagrange's equations are 
\be \label{Lag0}
\dot{q}=v, \quad \dot{p}=\frac{\partial L}{\partial q}, \quad
p=\frac{\partial L}{\partial v} .
\ee
We can introduce the extended Hamiltonian 
in the space of variables $(t,q,p,v)$ as
\be \label{eH}
H_{*}(t,q,p,v)=pv-L(t,q,v).
\ee
The corresponding extended Poincare-Cartan 1-form is defined by
\be \label{eCP}
\omega_{h*}=pdq-H^{*}dt=pdq+Ldt-pvdt .
\ee

{\bf Proposition 9.}\\
{\it The exterior derivative of the form (\ref{eCP})
is defined by the equation } \\
\be \label{deCP3}
d\omega_{h*}=[D_t p -D_q L] dt \wedge dq-
[dq - vdt ] \wedge dp-
[p -D_v L ]dv \wedge dt . \ee

\vskip 3mm

{\bf Proof.}
The exterior derivative of (\ref{eCP}) is
\[ d\omega_{h*}=d(pdq)-d(H^{*}dt)=
D_t p dt \wedge dq+D_q p dq \wedge dq+D_p p dp \wedge dq +D_v p dv \wedge dq- \]
\be \label{deCP1}
-D_t H^{*} dt \wedge dt- D_q H^{*} dq \wedge dt-D_p H^{*} dp \wedge dt
-D_v H^{*} dv \wedge dt . \ee
Using $dt\wedge dt=0$, $dq\wedge dt= -dt \wedge dq$, and
$D_q p=0$, $D_v p=0$, equation (\ref{deCP1}) gives
\be \label{deCP2}
d\omega_{h*}=
[D_t p +D_q H^{*}] dt \wedge dq-
[D_p p dq - D_p H^{*}dt ] \wedge dp-
D_v H^{*} dv \wedge dt . \ee
From (\ref{eH}), we get
\[ D_q H^{*}=D_q[pv-L]=-D_q L ,  \]
\[ D_p H^{*}=D_p [pv - L]=v D_p p-D_p L(t,q,v)=v D_p p=v , \]
\[ D_v H^{*}=D_v [pv - L]=pD_v v-D_v L=p-D_v L .\]
As the result, we obtain (\ref{deCP3}).

\vskip 3mm

{\bf Stationary Action Principle in Lagrange's Approach}  \\
{\it The trajectory of a dynamical system can be derived 
by finding the path for which the form (\ref{eCP})
is closed , i.e.,}
\be \label{eSAP}
d \omega_{h*}=0 .
\ee

From equations (\ref{deCP3}) and (\ref{eSAP}), we get 
\be \label{81}  D_t p -D_q L=0, \quad
dq - vdt=0 ,\quad p-D_v L=0 . \ee
It is easy to see that equation (\ref{81}) coincides with the Lagrange's  
equations (\ref{Lag0}) that can be presented as
\be
D_q L-\left[ \frac{d}{dt}  D_v L \right]_{v=\dot{q}}=0.
\ee 
As the result, we obtain 
\be \label{usualELE} 
\frac{\partial L}{\partial q_i}-
\frac{d}{dt} \left(\frac{\partial L}{\partial \dot{q}_i} \right)=0 , 
\quad i=1,..n ,
\ee
which are the Euler-Lagrange equations for Lagrangian system.

\subsection{Fractional Lagrange's Equations}

Suppose $L(t,q,v)$ is a Lagrangian of the system, and
the extended Hamiltonian is defined by 
\be \label{feH}
H_{*}(t,q,p,v)=pv^{\beta}-L(t,q,v) ,
\ee
where $\beta$ is a positive power.
Let us define 
\be \label{feCP}
\omega_{h*\alpha}=p(dq)^{\alpha}-H^{*}(dt)^{\alpha}=
p(dq)^{\alpha}+L(dt)^{\alpha}-pv^{\beta}(dt)^{\alpha} ,
\ee
which is a fractional generalization of
the extended Poincare-Cartan 1-form (\ref{eCP}). \\

{\bf Proposition 10.} \\
{\it The fractional exterior derivative of the fractional 1-form (\ref{feCP})
is defined by } \\
\[ d^{\alpha}\omega_{h*\alpha}=
\left[D^{\alpha}_t p -D^{\alpha}_q L \right] 
(dt)^{\alpha} \wedge (dq)^{\alpha}- \]
\be \label{dfeCP3}  -
(D^{\alpha}_p p)\left[ (dq)^{\alpha}- v^{\beta} (dt)^{\alpha} \right]
\wedge (dp)^{\alpha}
-\left[ pD^{\alpha}_v v^{\beta}-D^{\alpha}_v L \right] 
(dv)^{\alpha} \wedge (dt)^{\alpha} . 
\ee

\vskip 3mm

{\bf Proof.} 
The fractional exterior derivative of the form (\ref{feCP})
gives
\[ d^{\alpha}\omega_{h*\alpha}=
d^{\alpha}[pd^{\alpha}q]-d[H^{*}(dt)^{\alpha}]=
(D^{\alpha}_t p) (dt)^{\alpha} \wedge (dq)^{\alpha}+
(D^{\alpha}_q p) (dq)^{\alpha} \wedge (dq)^{\alpha}+ \]
\[ +(D^{\alpha}_p p) (dp)^{\alpha} \wedge (dq)^{\alpha}+
(D^{\alpha}_v p) (dv)^{\alpha} \wedge (dq)^{\alpha}- \]
\[ -(D^{\alpha}_t H^{*}) (dt)^{\alpha} \wedge (dt)^{\alpha}-
(D^{\alpha}_q H^{*}) (dq)^{\alpha} \wedge (dt)^{\alpha}-\]
\be \label{dfeCP1} -
(D^{\alpha}_p H^{*}) (dp)^{\alpha} \wedge (dt)^{\alpha}-
(D^{\alpha}_v H^{*}) (dv)^{\alpha} \wedge (dt)^{\alpha} . \ee
Using $(dt)^{\alpha}\wedge (dt)^{\alpha}=0$, 
$(dq)^{\alpha}\wedge (dt)^{\alpha}= -(dt)^{\alpha} \wedge (dq)^{\alpha}$, 
and $D_q p=0$, $D_v p=0$ for Riesz or Caputo fractional derivatives, 
we can rewrite (\ref{feCP}) as
\[ d^{\alpha}\omega_{h*\alpha}=
\left[D^{\alpha}_t p+D^{\alpha}_q H^{*} \right] 
(dt)^{\alpha} \wedge (dq)^{\alpha}- \]
\be \label{dfeCP2} 
-\left[(D^{\alpha}_p p) (dq)^{\alpha}-(D^{\alpha}_p H^{*}) (dt)^{\alpha} \right]
\wedge (dp)^{\alpha}
-(D^{\alpha}_v H^{*}) (dv)^{\alpha} \wedge (dt)^{\alpha} . \ee
Using the extended Hamiltonian (\ref{feH})
and properties of Riesz and Caputo derivatives, we have
\[ D^{\alpha}_q H^{*}=D^{\alpha}_q [pv^{\beta}- L]=-D^{\alpha}_q L ,  \]
\[ D^{\alpha}_p H^{*}=D^{\alpha}_p [pv^{\beta}- L]=
v^{\beta} D^{\alpha}_p p-D^{\alpha}_p L(t,q,v)=v^{\beta} D^{\alpha}_p p ,\]
\[ D^{\alpha}_v H^{*}=D^{\alpha}_v [pv^{\beta} - L]=
pD^{\alpha}_v v^{\beta}-D^{\alpha}_v L .\]
As the result, we obtain (\ref{dfeCP3}). \\

{\bf Fractional Action Principle in Lagrange's Approach}  \\
{\it The trajectory of fractional dynamical systems can be derived 
by finding the path for which the 
fractional extended Poincare-Cartan 1-form (\ref{feCP})
is closed form, i.e.,} 
\be \label{feSAP}
d^{\alpha} \omega_{h*\alpha}=0 .
\ee

Using (\ref{dfeCP3}) and (\ref{feSAP}), we have 
\be 
(D^{\alpha}_t p) -D^{\alpha}_q L =0, \quad
(dq)^{\alpha}- v^{\beta} (dt)^{\alpha}=0 ,\quad
pD^{\alpha}_v v^{\beta}-D^{\alpha}_v L =0 . 
\ee
From the relation
\be
D^{\alpha}_v v^{\beta}=
\frac{\Gamma(\beta+1)}{\Gamma(\beta+1-\alpha)} v^{\beta-\alpha},
\ee
where $\beta>-1$, we get
\be \label{fLagr1}
D^{\alpha}_t p=D^{\alpha}_q L, \quad
v^{\beta}=\frac{(dq)^{\alpha}}{(dt)^{\alpha}} , \quad
p= \frac{\Gamma(\beta+1-\alpha)}{\Gamma(\beta+1)}
v^{\alpha-\beta} D^{\alpha}_v L . 
\ee
Substitution of  the third equation from (\ref{fLagr1}) into 
the first one gives 
\be \label{fLagr2}
D^{\alpha}_q L-\frac{\Gamma(\beta+1-\alpha)}{\Gamma(\beta+1)}
D^{\alpha}_t \left[ v^{\alpha-\beta}D^{\alpha}_v L  
\right]_{v^{\beta}=(\dot{q})^{\alpha}}=0 .
\ee 
Equation (\ref{fLagr2}) has the dependence
\be v^{\beta}=(\dot{q})^{\alpha} . \ee
It is easy to see that equation (\ref{fLagr2}) looks unusually 
even for $\beta=1$.
Therefore we use $\beta=\alpha$
for the Hamiltonian (\ref{feH}) and the form (\ref{feCP}). 


Using $\beta=\alpha$ in equations (\ref{feH}), (\ref{feCP}), and
\be
D^{\alpha}_v v^{\alpha}=\frac{1}{\Gamma(2-\alpha)} ,
\ee
we have the fractional extended Lagrange's equations
\be \label{fLagr3}
(D^{\alpha}_t p)=D^{\alpha}_q L, \quad
v^{\alpha}=\frac{(dq)^{\alpha}}{(dt)^{\alpha}} , \quad
p= \Gamma(2-\alpha) D^{\alpha}_v L . 
\ee
Substituting the third equation from (\ref{fLagr4}) into the first one, 
we  obtain 
\be \label{fLagr4}
D^{\alpha}_q L-\Gamma(2-\alpha)
D^{\alpha}_t \left[ D^{\alpha}_v L  \right]_{v=\dot{q}}=0 ,
\ee 
that is {\it fractional Euler-Lagrange equations}.
As the result, the fractional equations of motion for Lagrangian systems 
are presented by
\be \label{EL}
\hat E^{\alpha}_i L(t,q,\dot{q})=0, \quad i=1,...,n,
\ee
where
\be
\hat E^{\alpha}_i=
D^{\alpha}_{q_i}-\Gamma(2-\alpha) D^{\alpha}_{t} D^{\alpha}_{\dot{q}_{i}} .
\ee
For $\alpha=1$, equations (\ref{EL}) are 
the Euler-Lagrange equations (\ref{usualELE}).\\

\section{Fractional Non-Hamiltonian Systems}

\subsection{Helmholtz Conditions}

In this subsection, we consider the brief review of
Helmholtz conditions \cite{Helm,FSS} to fix notation and provide 
a convenient reference.
The Helmholtz equations give the necessary and sufficient conditions 
for equations to be the Euler-Lagrange equations
that can be derived from stationary action principle. \\

{\bf Proposition 11.}
{\it The necessary and sufficient conditions for 
\be \label{Fm}
F_{i}(t,q,\dot{q},...,q^{(N)})=0, \quad i=1,...,n
\ee
to be equations 
that can be derived from the stationary action principle are
\be \label{P1}
\frac{\partial F_{i}}{\partial q_{j}}-
\frac{\partial F_{j}}{\partial q_{i}}-
\sum^{N}_{k=1} (-1)^k \frac{d^k}{dt^k} 
\left(\frac{\partial F_{j}}{\partial q^{(k)}_{i}}\right)=0,
\ee
\be\label{P2}
\frac{\partial F_{i}}{\partial q^{(m)}_{j}}-
\sum^{N}_{k=m} (-1)^k (^k_m) \frac{d^{k-m}}{dt^{k-m}} 
\left(\frac{\partial F_{j}}{\partial q^{(k)}_{i}}\right)=0,
\quad m=1,...,N
\ee
where $q^{(k)}_{i}=d^kq_{i}/dt^k$, $i,j=1,...n$ and} \\
\be
(^k_m)=\frac{k!}{m!(k-m)!}, \quad
\frac{d}{dt}=\frac{\partial}{\partial t}+\sum^N_{k=1} q^{(k)}_{i}
\frac{\partial^k}{\partial q^{(k-1)}_{i}} .
\ee

{\bf Proof}. This proposition is proved in \cite{FSS}. \\

For simple example, let us consider the equations
\be
F_{i}(t,q,\dot{q})=0, 
\ee
Conditions (\ref{P1}) and (\ref{P2}) have the form
\be \label{P3}
\frac{\partial F_{i}}{\partial q_{j}}-
\frac{\partial F_{j}}{\partial q_{i}}+
\frac{d}{dt} 
\left(\frac{\partial F_{j}}{\partial \dot{q}_{i}}\right)=0,
\ee
\be\label{P4}
\frac{\partial F_{i}}{\partial \dot{q}_{j}}+
\frac{\partial F_{j}}{\partial \dot{q}_{i}}=0,
\quad i,j=1,...,n,
\ee
where 
\be
\frac{d}{dt}=\frac{\partial}{\partial t}+\dot{q}_{i} 
\frac{\partial}{\partial q_{i}} .
\ee
Equations (\ref{P4}) gives
\be \label{P5}
\frac{\partial^2 F_{i}}{\partial \dot{q}_{j} \partial \dot{q}_{\kappa}}=0 ,
\quad i,j,\kappa=1,...,n.
\ee
These conditions are satisfied for the linear
dependence $F_{i}$ with respect to $\dot{q}$, i.e.,
\be \label{P6}
F_{i}=C_{ij}(t,q) \dot{q}_{j} +D_{i}(t,q)=0, \quad 
i=1,...,n.
\ee

{\bf Corollary 1.}
{\it The necessary and sufficient conditions to derive 
equations (\ref{P6}) from the stationary action principle
have the form}
\be \label{P7}
C_{ij}=-C_{ji},
\ee
\be \label{P8}
\frac{\partial C_{ij}}{\partial q_{\kappa}}+
\frac{\partial C_{j \kappa}}{\partial q_{i}}+
\frac{\partial C_{\kappa i}}{\partial q_{j}}=0,
\ee
\be \label{P9}
\frac{\partial C_{ij}}{\partial t}-
\frac{\partial D_{i}}{\partial q_{j}}+
\frac{\partial D_{j}}{\partial q_{i}}=0.
\ee

{\bf Proof.}
Substitution of equation (\ref{P6}) into 
equations (\ref{P3}) and (\ref{P4}) yelds
(\ref{P7}), (\ref{P8}), and (\ref{P9}). \\

{\bf Corollary 2.}
{\it The necessary and sufficient conditions for 
\be 
F_{i}(t,q,\dot{q},\ddot{q})=0, \quad i=1,...,n 
\ee
to be equations that can be derived from stationary action principle are
\be \label{P10}
\frac{\partial F_{i}}{\partial \ddot{q}_{j}}-
\frac{\partial F_{j}}{\partial \ddot{q}_{i}}=0, 
\ee
\be \label{P11}
\frac{\partial F_{i}}{\partial \dot{q}_{j}}+
\frac{\partial F_{j}}{\partial \dot{q}_{i}}-2
\frac{d}{dt} 
\left(\frac{\partial F_{j}}{\partial \ddot{q}_{i}}\right)=0,
\ee
\be \label{P12}
\frac{\partial F_{i}}{\partial q_{j}}-
\frac{\partial F_{j}}{\partial q_{i}}+
\frac{d}{dt} 
\left(\frac{\partial F_{j}}{\partial \dot{q}_{i}}\right)
-\frac{d^2}{dt^2} 
\left(\frac{\partial F_{j}}{\partial \ddot{q}_{i}}\right)=0,
\ee
where }
\be
\frac{d}{dt}=\frac{\partial}{\partial t}+\dot{q}_{i} 
\frac{\partial}{\partial q_{i}}+\ddot{q}_{i} 
\frac{\partial}{\partial \dot{q}_{i}} .
\ee

Note that using equation (\ref{P11}) condition (\ref{P12}) 
can be rewritten in the more symmetric form
\be \label{P12b}
\frac{\partial F_{i}}{\partial q_{j}}-
\frac{\partial F_{j}}{\partial q_{i}}-
\frac{1}{2}\frac{d}{dt} \left(
\frac{\partial F_{i}}{\partial \dot{q}_{j}}-
\frac{\partial F_{j}}{\partial \dot{q}_{i}}\right)=0.
\ee

\subsection{Non-Lagrangian Systems}

{\bf Definition 5.}
{\it A dynamical system is called non-Lagrangian system
if the equations of motion (\ref{Fm}) cannot be represented
in the form
\be 
\sum^N_{k=0} (-1)^{k} \frac{d^k}{dt^k}  
\frac{\partial^k}{\partial q^{(k)}} L(t,q,\dot{q},...,q^{(N)})=0 ,
\ee
with some function $L=L(t,q,\dot{q},...,q^{(N)})$, 
where $q^{(k)}=d^k q/dt^k$.} 

It is well-known that the equations 
of second order cannot be presented as
\be
\hat E_i L(t,q,\dot{q})=0 , \quad i=1,...,n,
\ee
where $\hat E_i$ is the Euler-Lagrange operator 
\be
\hat E_i=D_{q_i}-D_t D_{\dot{q}_i}.
\ee
In the general case, the Lagrange's equations 
have the additional term $Q_i(t,q,\dot{q})$
which is a generalized non-potential force.
This force cannot be presented as $Q_i=\hat E_i U$
for some function $U=U(t,q,\dot{q})$. 
In general, the Euler-Lagrange equations  \cite{Gold} are 
\be
\hat E_i L(t,q,\dot{q})+Q_i=0.
\ee

If we consider non-potential forces and non-Lagrangian systems, 
then the nonholonomic variational 
equation suggested by L.I. Sedov \cite{Sedov1,Sedov2,Sedov3,Sedov4}
should be used instead of stationary action principle.

\subsection{Non-Hamiltonian Systems and Friction Force}

In general, the phase-space equations of  motion 
cannot be presented in the form
\be \label{Hs}
\dot{q}_i=D_{p_i} H, \quad \dot{p}_i=-D_{q_i} H,
\ee
where $H=H(t,q,p)$ is a smooth function. 
The Hamilton's equations are written as
\be \label{nHs}
\dot{q}_i=D_{p_i} H+G^i(t,q,p), \quad \dot{p}_i=-D_{q_i} H+F^i(t,q,p),
\ee
where $H=H(t,q,p)$ is a Hamiltonian of the system.
For example, $H(t,q,p)=T(p)+U(q)$,
where $T(p)$ is a kinetic energy, and $U(q)$ is potential energy 
of the system.
The functions $G^i(t,q,p)$ and $F^i(t,q,p)$ describe 
the non-potential forces which act on the system. 
For mechanical systems, we can consider $G^i(t,q,p)=0$.
If the functions $G^i(t,q,p)$ and $F^i(t,q,p)$
do not satisfy the Helmholtz conditions (\ref{HC1}), then 
(\ref{nHs}) is a non-Hamiltonian system. 

In general, the exterior derivative of the Poincare-Cartan 1-form
is not equal to zero ($d\omega_h\not=0$).  
This derivative is equal to differential 2-form $\theta$
that is defined by non-potential forces 
\be \label{theta}
\theta=F^i(t,q,p) dt \wedge dq_i-G^i(t,q,p) dt \wedge dp_i 
\ee
for the non-Hamiltonian system (\ref{nHs}).
For example, the linear friction force $F^i=-\gamma p_i$ gives
\be
\theta=-\gamma p_i dt \wedge dq_i .
\ee

\vskip 3mm

{\bf Proposition 12.}
{\it The differential 2-form $\theta$ of non-potential forces
is non-closed form.} \\

{\bf Proof.}
If differential 2-form $\theta$ is a closed form ($d\theta=0$)
on a contractible open subset X of $R^{2n}$, then the 
form is the exact form such that a function $h=h(t,q,p)$ exists,
and $\theta=d h$. In this case, we have a new Poincare-Cartan 1-form
\[ \omega^{\prime}_h=\omega_h+h, \] 
such that $d \omega^{\prime}=0$, and the system is Hamiltonian. \\

There is a generalization of stationary action 
principle for the systems with non-potential forces. \\

{\bf Action Principle for non-Hamiltonian Systems}  \\
{\it The trajectory of a non-Hamiltonian system can be derived 
by finding the path for which the exterior derivative 
of the action 1-form (\ref{CP}) is equal to the
non-closed 2-form (\ref{theta}), i.e.,  }
\be \label{Sed1}
d \omega_h=\theta .
\ee

Equations (\ref{dCP3}), (\ref{theta}) and (\ref{Sed1}) give 
the equations of motion (\ref{nHs}) 
for non-Hamiltonian system.

\subsection{Fractional Generalization of non-Hamiltonian Systems}

Let us define a fractional generalization of the form (\ref{theta}) by
\be \label{ftheta}
\theta_{\alpha}=
F^i(t,q,p) (dt)^{\alpha} \wedge (dq_i)^{\alpha}-
G^i(t,q,p) (dt)^{\alpha} \wedge (dp_i)^{\alpha} .
\ee
This form allows us to derive fractional equations of motion
for non-Hamiltonian systems. \\

{\bf Fractional Action Principle for non-Hamiltonian Systems}  \\
{\it The trajectory of a fractional system subjected by non-potential 
forces can be derived 
by finding the path for which the fractional exterior derivative 
of the fractional action 1-form (\ref{fCP}) is equal to 
non-closed fractional 2-form (\ref{ftheta}), i.e. }
\be \label{Sed2}
d^{\alpha} \omega_{h\alpha}=\theta_{\alpha} .
\ee

Using (\ref{dfCP3}), (\ref{ftheta}) and (\ref{Sed2}), we get 
\be 
\frac{p^{1-\alpha}}{\Gamma(2-\alpha)} (dq)^{\alpha}-
D^{\alpha}_p H (dt)^{\alpha}=G(t,q,p) (dt)^{\alpha}, \quad 
D^{\alpha}_t p=-D^{\alpha}_q H +F(t,q,p).
\ee
As the result, we obtain
\be 
\left(\frac{d q_i}{dt}\right)^{\alpha}=\Gamma(2-\alpha)
p^{\alpha-1}_i D^{\alpha}_{p_i} H+ G^i(t,q,p),\quad
D^{\alpha}_t p_i=-D^{\alpha}_{q_i} H+F^i(t,q,p) .
\ee
These equations are the fractional generalization
of equations of motion for non-Hamiltonian systems.

\newpage
\section{Conclusion}

In this paper, we define a fractional exterior derivative 
for calculus of variations.
Hamiltonian and Lagrangian approaches are considered.
Hamilton's and Lagrange's equations with fractional derivatives
are derived from the stationary action principles.
We prove that fractional equations can be derived from
action which has only integer derivatives.
Derivatives of noninteger order appear by the fractional variation of 
Lagrangian and Hamiltonian.

Application of fractional variational calculus 
can be connected with a generalization of variational problems.
The gradient systems (GS) form a restricted class of 
ordinary differential equations.
Equations for GS can be defined by one function - potential.
Therefore the study of GS can be reduced to research of potential.
As a physical example, the ways of some chemical reactions 
are defined from the analysis of potential energy surfaces \cite{LB,K}.
The fractional gradient systems (FGS)
have been suggested in \cite{JPA2005-2}.
It was proved that GS are a special case of such systems.  
FGS includes a wide class of non-gradient systems. 
For example, the Lorenz and Rossler equations
are fractional gradient systems \cite{JPA2005-2}.
Therefore the study of the non-gradient system 
which are FGS can be reduced to research of potential.

We can assume that the ways of some chemical reactions with 
dissipation and systems with deterministic chaos can be considered 
by the analysis of fractional potential surfaces.
Let us note the interesting property of potential 
surfaces for systems with strange attractors.
The surfaces of the stationary states of the Lorenz and 
Rossler equations separate the three-dimensional Euclidean space 
into some number of areas \cite{JPA2005-2}. 
We have eight areas for the Lorenz equations and
four areas for the Rossler equations.
This separation has the interesting property:
all regions are connected with each other \cite{JPA2005-2}. 
Beginning movement from one of the areas, it is possible 
to appear in any other area, not crossing a surface.
Any two points from different areas can be connected 
by a curve which does not cross a surface.

The fractional variations can be used to define
the fractional generalization of gradient type equations 
that have the wide application for 
the description dissipative structures \cite{Prig,ZaslavskyBook2}.
The fractional gradient type equations are  
generalization of FGS \cite{JPA2005-2}
from ordinary differential equations into
partial differential equations. 
We plan to realize this generalization in the next paper by using
de Donder--Weyl Hamiltonian and Poincare-Cartan n-form.


\end{document}